\documentclass[5p]{elsarticle}

\usepackage{hyperref}

\usepackage{amsmath}
\usepackage{upgreek}

\newcommand{\Neq}{\ensuremath{\text{n}_{\text{eq}}\,\text{cm}^{-2}}}
\newcommand{\irniel}[2]{\ensuremath{#1\cdot10^{#2}\,\Neq}}

\journal{Nuclear Instruments and Methods in Physics Research Section A}

\begin{document}

\begin{frontmatter}

\title{Active Pixel Sensors in ams H18/H35 HV-CMOS Technology for the ATLAS HL-LHC Upgrade}

\author[mymainaddress,mysecondaryaddress]{Branislav Ristic}
\ead{bristic@cern.ch}

\author{for the ATLAS CMOS Pixel Collaboration}

\address[mymainaddress]{CERN, 1211 Gen\`{e}ve 23, Switzerland}
\address[mysecondaryaddress]{DPNC, Universit\'{e} de Gen\`{e}ve, Quai Ernest-Ansermet 24, 1211 Gen\`{e}ve 4, Switzerland}

\begin{abstract}
Deep sub micron HV-CMOS processes offer the opportunity for sensors built by industry standard techniques while being HV tolerant, making them good candidates for drift-based, fast collecting, thus radiation-hard pixel detectors.
For the upgrade of the ATLAS Pixel Detector towards the HL-LHC requirements, active pixel sensors in HV-CMOS technology were investigated.
These implement amplifier and discriminator stages directly in insulating deep n-wells, which also act as collecting electrodes.
The deep n-wells allow for bias voltages up to 150\,V leading to a depletion depth of several 10$\,\upmu$m.
Prototype sensors in the ams H18 180\,nm and H35 350\,nm HV-CMOS processes have been manufactured, to assess a potential drop-in replacement for the current ATLAS Pixel sensors, thus leaving higher level processing such as trigger handling to dedicated read-out chips.

Sensors were thoroughly tested in lab measurements as well as in testbeam experiments.
Irradiations with X-rays and protons revealed a tolerance to ionizing doses of $1\,$Grad.
An enlarged depletion zone of up to 100$\,\upmu$m thickness after irradiation due to the acceptor removal effect was deduced from Edge-TCT studies.
Finally, the sensors showed high detection efficiencies after neutron irradiation to $10^{15}\,\Neq$ in testbeam experiments.

A full reticle size demonstrator chip, implemented in the H35 process is being submitted to prove the large scale feasibility of the HV-CMOS concept.
\end{abstract}


\end{frontmatter}


\section{HV-CMOS sensors for the ATLAS Inner Tracker}
\begin{figure}[htbp]
    \centering
    \includegraphics[width=.8\linewidth]{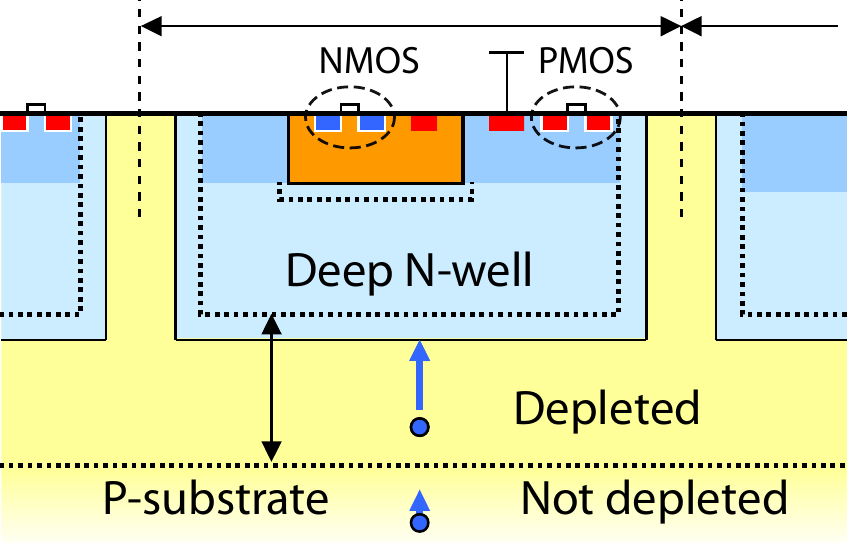}
    \caption{Simplified cross section of an HV-CMOS sensor. The highly sensitive depletion zone grows from the deep n-well implants that shield the electronics sitting on the surface of the chip from the high voltage in the bulk}
    \label{fig:hvcmos}
  \end{figure}

The sensors of the ATLAS\footnote{A Toroidal LHC ApparatuS} Inner Detector, especially the Pixel Detector are located closest to the interaction point, thus experiencing the highest density of particles and radiation levels.
After the Phase II upgrade, the LHC\footnote{Large Hadron Collider} is expected to deliver an integrated luminosity of up to $3000\,\text{fb}^{-1}$ to the ATLAS Detector inflicting a total ionizing dose beyond $1\,$Grad and NIEL\footnote{Non Ionizing Energy Loss} fluences beyond $10^{16}\,\Neq$ on the innermost Pixel layer \cite{atlasloi}.
In order to cope with these elevated requirements the Inner Detector will be replaced by a purely silicon based one.
For the inner layers high granularity combined with an extremely radiation hard design is necessary while going to larger radii the cost of producing large areas becomes the limiting factor for detector construction.\\
The High-Voltage CMOS\footnote{Complementary Metal-Oxide-Semiconductor} technology offers a new approach of producing silicon sensors in a process, standardized and widely used by industry for high voltage switching electronics.
Deep n-wells implanted in a p-substrate (see figure \ref{fig:hvcmos}) provide the detecting pn-junctions, thus acting as collecting electrodes while shielding the electronics on the surface of the sensor from the applied high voltage.
The highly doped substrate is intrinsically radiation hard, but provides only shallow depletion zones.
For a good SNR\footnote{Signal to Noise Ratio}, amplification and further electronics are implemented inside the deep n-well.

Current pixel sensors are usually passive and need a dedicated readout chip for signal processing to which they are bump bonded.
This process is usually expensive due to the small pixel pitch of $50\,\upmu$m, but can be avoided by coupling capacitively via gluing.
Although in principle possible for passive sensors, the necessary signal size is difficult to reach without on-sensor amplification, as provided by HV-CMOS sensors.

\section{Active pixel sensors in the ams 180nm High-Voltage CMOS technology} 
The prototypes considered in this paper were built in the Austria MicroSystems (ams) 180nm HV-CMOS process H18 based on designs by I. Peric\cite{hvcmos}.
Deep n-well structures implanted in a low resistivity ($\sim 10\, \Omega \text{cm}$) p-type substrate allow for a bias voltage of up to $95\,$V which is expected to yield a depletion depth in the order of $10-20\,\upmu$m.
MIP\footnote{Minimum Ionizing Particle}-like particles traversing through that zone would deposit charge with an MPV\footnote{Most Probable Value} of around $600\ldots1200$ electron hole pairs.

\begin{figure}[htbp]
    \centering
    \includegraphics[width=.8\linewidth]{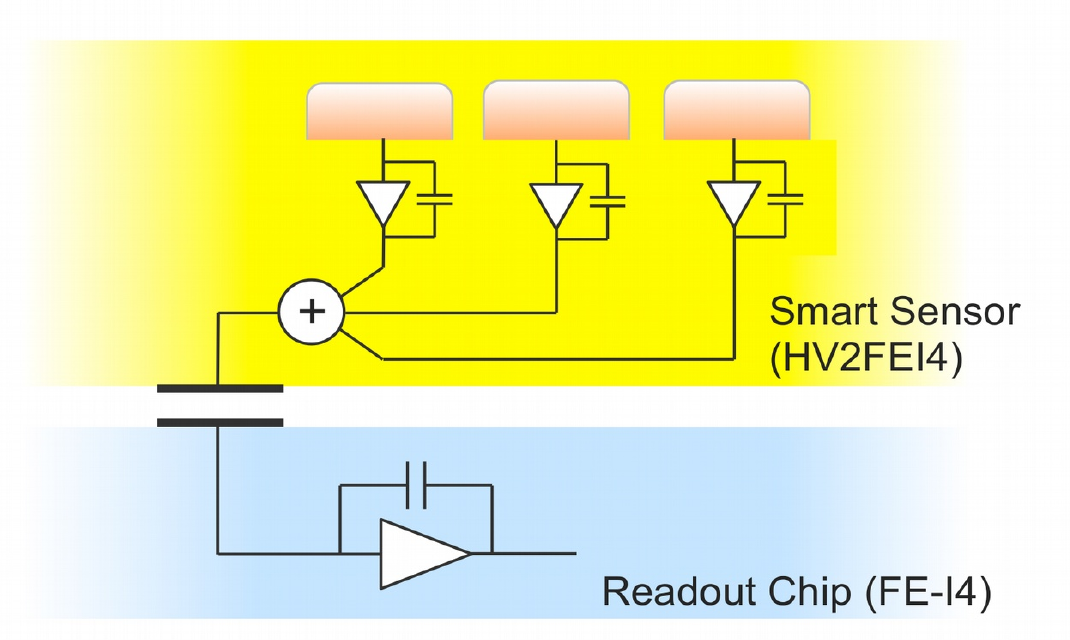}
    \caption{Sketch of an HV-CMOS sensor, capacitively coupled to the Frontend. Three sub pixels are connected additively to one Frontend cell.}
    \label{fig:aps_hybrid}
  \end{figure}

The sensors were designed for the latest ATLAS pixel readout chip FE-I4\footnote{Frontend-I4}\cite{fei4} implementing multiple pixel flavors with different optimizations concerning radiation hardness, low noise levels and speed.
The default pixel cell combines a charge sensitive amplifier and a source follower with a two stage discriminator.
Threshold levels can be set globally with local corrections, applied by setting per pixel DACs\footnote{Digital Analog Converter}.
The H18 CCPDv4\footnote{Capacitively Coupled Pixel Detector}, which will be discussed in this paper, consists mainly of $125\times33\,\upmu\text{m}^2$ pixels of which triples are connected capacitively to one Frontend pad in order to ensure the necessary footprint, as depicted in figure \ref{fig:aps_hybrid}.
Two Frontend cells, consisting of 6 HV-CMOS sub pixels, form one unit cell.
A sub pixel position encoding in the unit cell can be achieved by setting different output levels for the three sub pixels of one Frontend cell and later be decoded by the Frontend's analog readout.
Although in principle possible the encoding was not used for this paper.

Prototypes were irradiated with X-rays and protons at CERN and thermal neutrons at the TRIGA reactor of JSI Ljubljana, Slovenia\cite{ljubljana}.

\section{Irradiation with ionizing particles}
Ionizing radiation damages primarily the electronics implanted in the deep n-wells of the HV-CMOS sensors.
In order to assess this influence on the ams H18 sensors, prototypes have been irradiated up to $1\,$Grad TID\footnote{Total Ionizing Dose} with X-rays and up to $962\,$Mrad TID with $24\,$GeV protons.
Irradiations were done in multiple steps, between which the properties of the two types of amplifiers, with linear (LT) or enclosed feedback transistor (ELT), were monitored.

\begin{figure}[htbp]
    \centering
    \includegraphics[width=\linewidth]{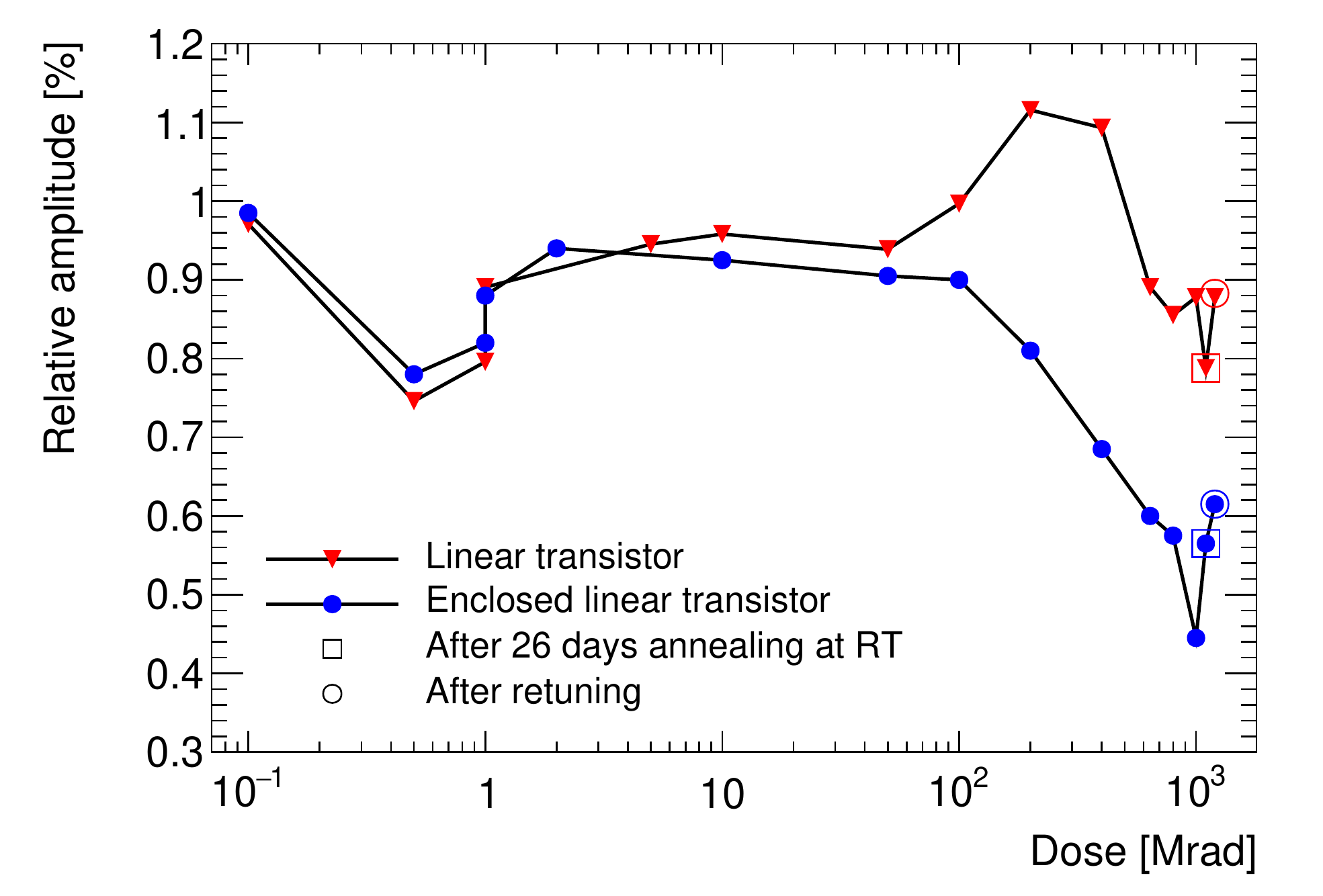}
    \caption{Response of two types of amplifiers, with linear and enclosed feedback transistors to an $1.2\,$V injection pulse after X-ray irradiation up to $1\,$Grad. The last three points were taken after $1\,$Grad, but separated for better visibility}
    \label{fig:irrad_xray_ampl}
  \end{figure}

The amplifier output after X-ray irradiation can be found in figure \ref{fig:irrad_xray_ampl}.
Both amplifiers were functional after $1\,$Grad, however with a resulting relative amplitude of $88\,\%$ for the LT and $45\,\%$ for the ELT flavor.
Although the irradiated LT showed a higher amplitude it became significantly noisy, while the ELT retained a clean signal.
The performance of the ELT amplifier recovered after $25$ days of annealing at room temperature with a final amplitude of $62\,\%$ after retuning of DAC parameters.
Room temperature annealing of the LT amplifier worsened the performance, but was also mitigated by retuning.

\begin{figure}[htbp]
    \centering
    \includegraphics[width=\linewidth]{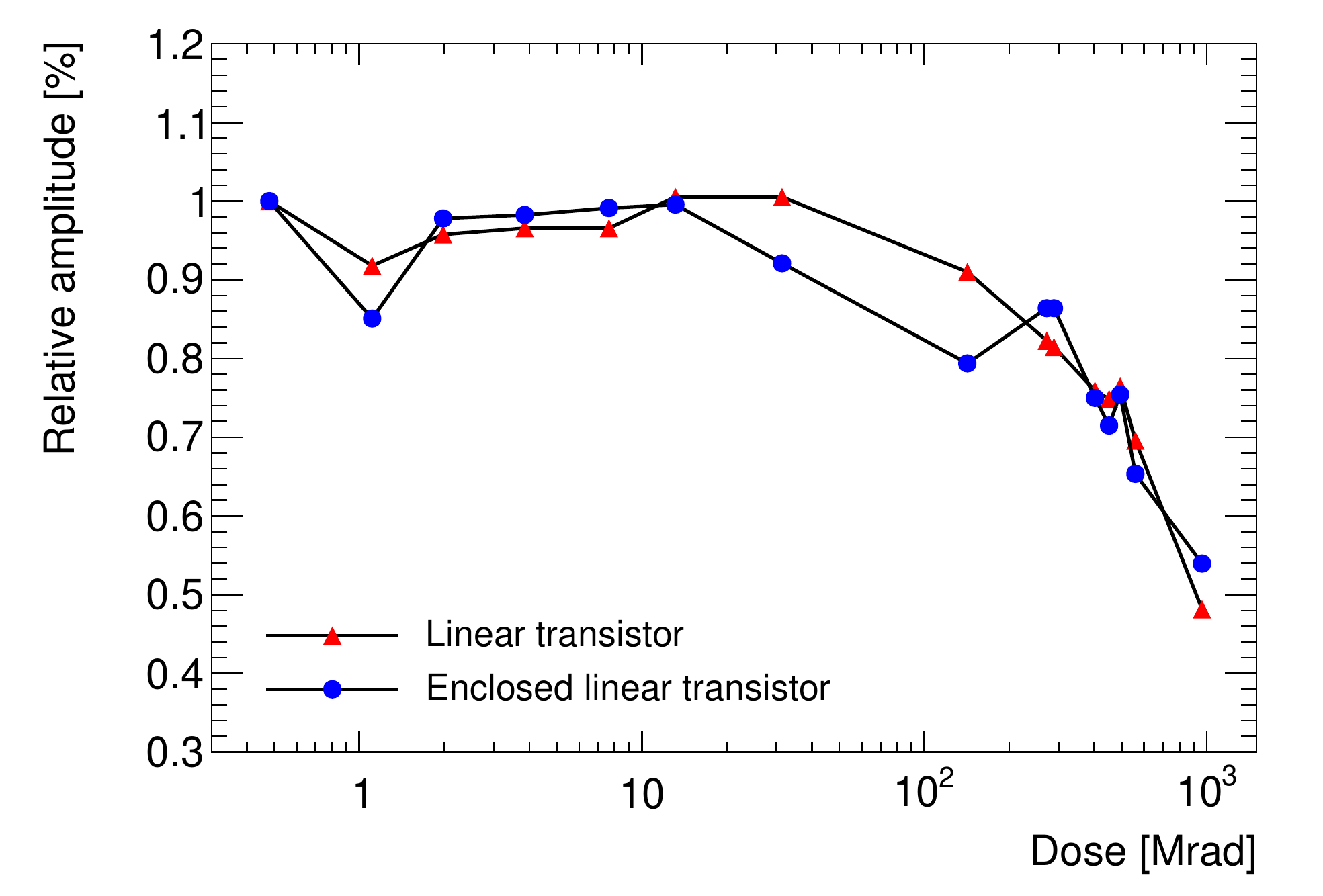}
    \caption{Response of two types of amplifiers, with linear and enclosed feedback transistors to an $1.2\,$V injection pulse after irradiation with protons up to $962\,$Mrad}
    \label{fig:irrad_proton_ampl}
  \end{figure}

When irradiated by protons the performance of both amplifiers is slightly increased between $2$ and $50\,$Mrad and then falls to $75\,\%$ as seen in figure \ref{fig:irrad_proton_ampl}.
As well as for X-rays, the LT becomes noisy after proton irradiation.
For this study neither annealing nor retuning was performed.
An increase in amplitude after retuning is conceivable.
  
\section{Testbeam results}
Prototypes of the ams H18 CCPDv4 design were characterized in a testbeam campaign at the CERN SPS\footnote{Super Proton Synchrotron} H8 $180\,$GeV pion beamline using the FE-I4 beam telescope of University of Geneva.
The measurements focused on the STime pixels, that incorporate the newest pixel design.
Although in principle possible, no sub pixel encoding was implemented in the DAQ\footnote{Data AcQuisition}, therefore all sub pixel outputs were set to the same value for a homogeneous signal in the Frontend.
This also warranted to merge every pair of FE pixels that corresponds to a unit cell forming virtual pixels of $250\times100\,\upmu\text{m}^2$ size.  

\begin{figure}[htbp]
    \centering
    \includegraphics[width=\linewidth]{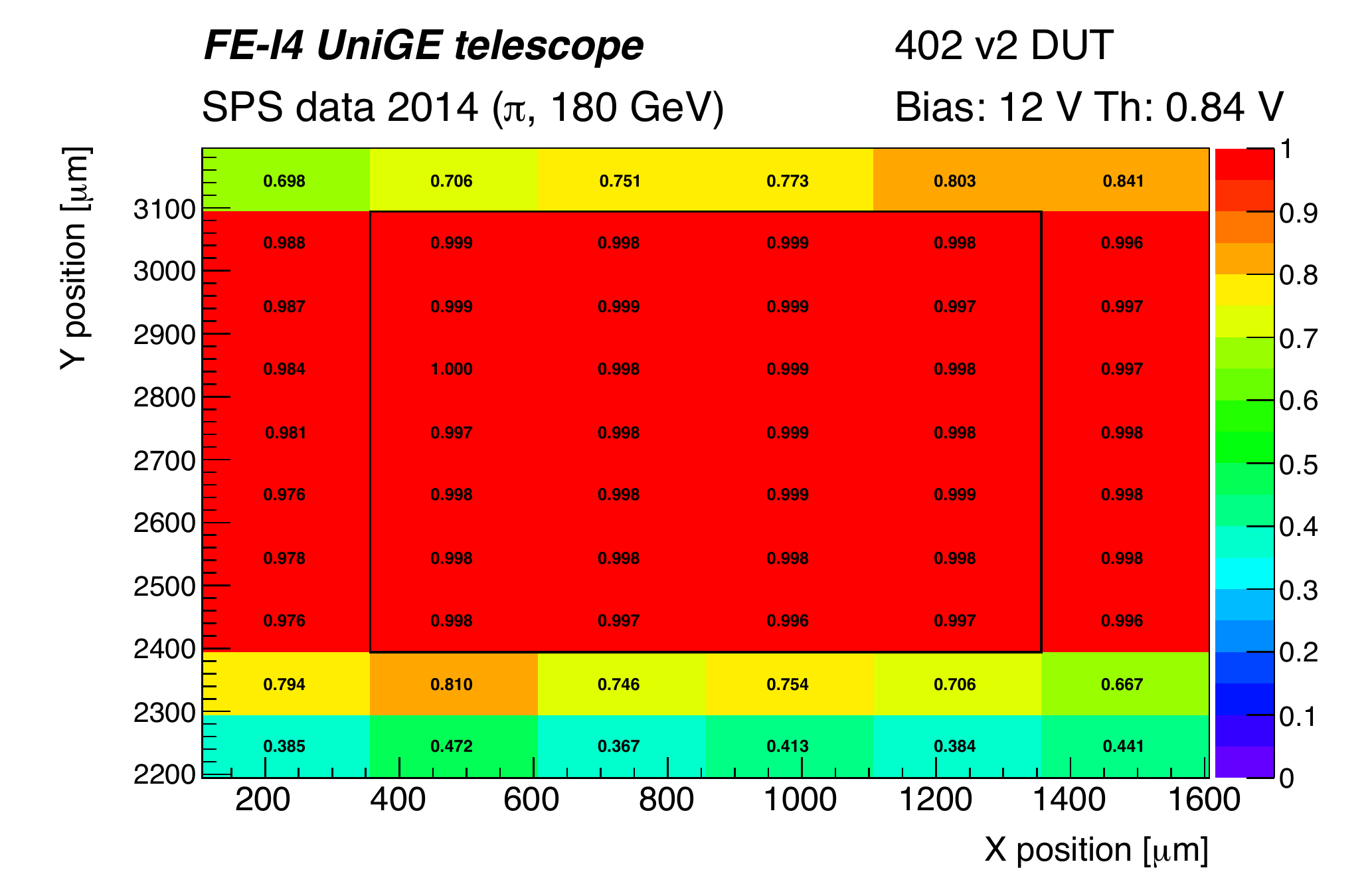}
    \caption{Efficiency map of the non irradiated sample 402. The lower low-efficiency band comes from an untuned pixel flavor while the upper one is a reconstruction artifact. Therefore only the matrix enclosed by the black box was considered for efficiency analysis}
    \label{fig:tb_Eff_402}
  \end{figure}

Tests were performed on the non irradiated sample 402 and the sample 404 which was irradiated to \irniel{1}{15}.
Due to early breakdown, the maximum voltage was limited to $12\,$V for 402 and $35\,$V for 404 instead of $95\,$V given by design.
Discriminators on both sensors were tuned to a threshold close to the noise edge, resulting in an MPV of the threshold of \mbox{$\approx 300\ldots500\,$}e.
The efficiency vs. virtual pixel position for 402 at its maximum bias voltage is depicted in figure \ref{fig:tb_Eff_402}.
As the measurements focused on the STime pixels, other pixel flavors, implemented on the sensor, were not tuned which results in the low efficiency band at the bottom part of the plot.
The upper low efficiency band is the result of a data reconstruction artifact and must not be considered further.
For both sensors a bias voltage scan has been performed and the efficiency computed, excluding the discussed artifacts.
The non irradiated 402 shows a high efficiency of over $98\,\%$ already without applied high voltage which increases to $99.7\,\%$ at $13\,$V.
Sample 404's efficiency raises from $\approx30\,\%$ in the unbiased case to $96.2\,\%$  at $35\,$V.

This indicates strongly that a big fraction of the charge is collected by diffusion, as after \irniel{1}{15} most of the diffusing charge carriers will be trapped, thus not contributing to the signal.
By applying a high bias voltage the sensor's performance can be restored. 
However this could not be tested fully, due to the early breakdown.
A significant increase of efficiency can be expected for a properly biased sensor as the data for 404 suggests.
  \begin{figure}[htbp]
    \centering
    \includegraphics[width=.95\linewidth]{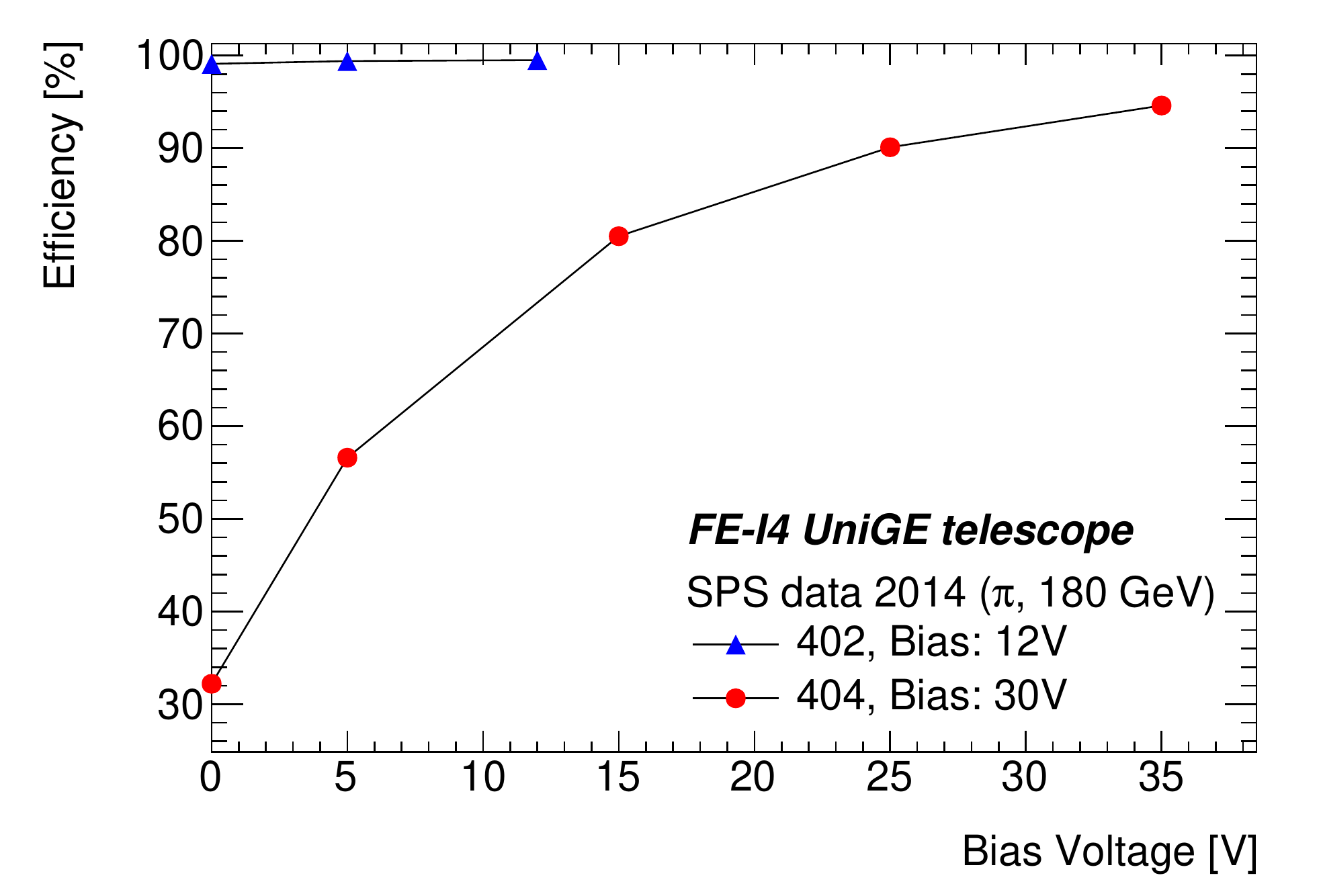}
    \caption{Efficiency with respect to applied bias voltage for 402 and 404 at the highest possible bias voltages}
    \label{fig:tb_EffVsBias_402_404}
  \end{figure}
  
\begin{figure}[htbp]
    \centering
    \includegraphics[width=.95\linewidth]{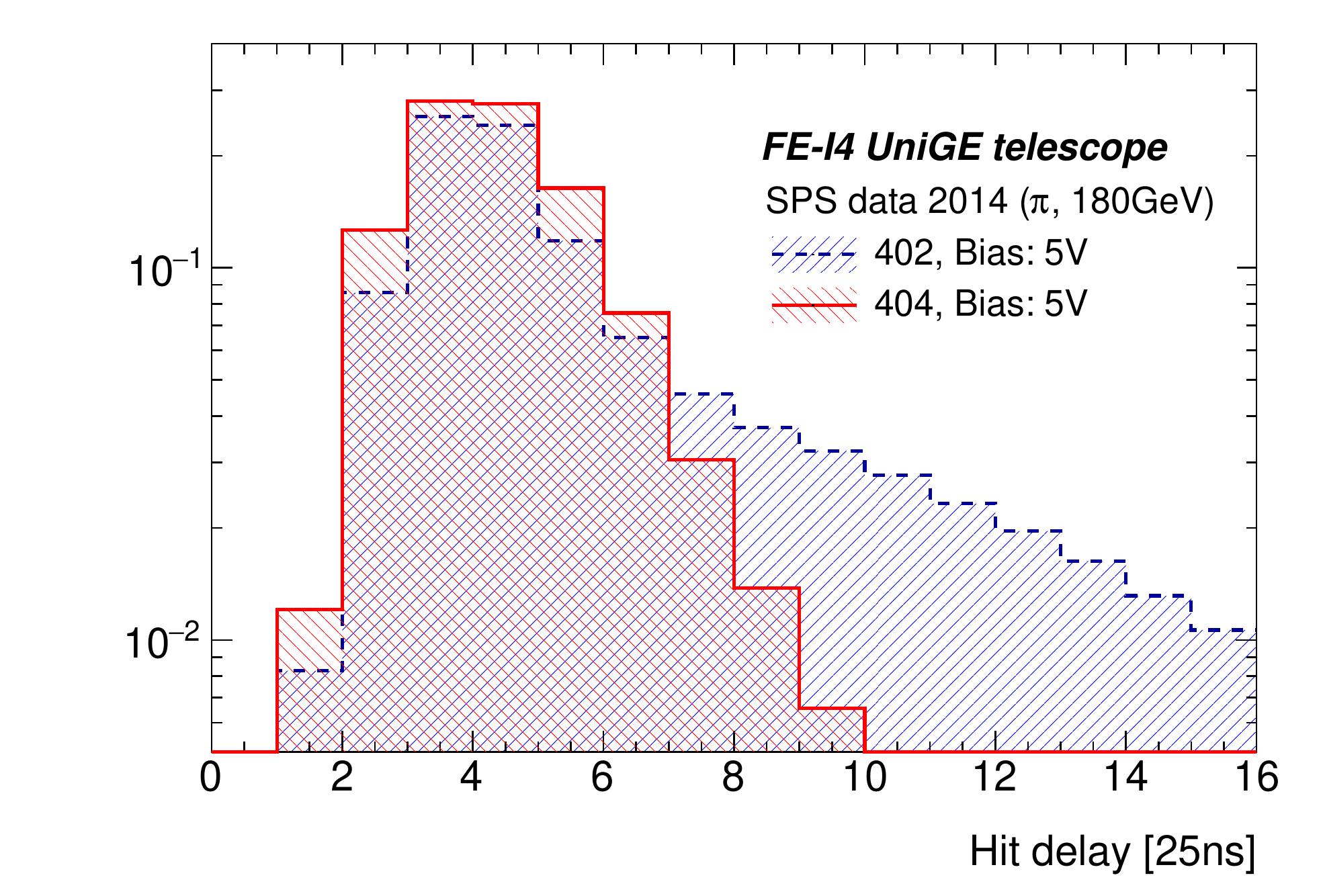}
    \caption{Distribution of the delay time between the telescope trigger and detection in the sensor in units of FE-I4 timing bins [25ns].}
    \label{fig:tb_lv1distr}
  \end{figure}

The assumption that diffusion is responsible for a big part of the signal is further supported by timing measurements.
Figure \ref{fig:tb_lv1distr} shows the distribution of the delay between the detection of the impacting particle by the telescope and the detection by the sensors biased at $5\,$V.
The uncertainty of this measurement is in the order of $20\,$ns.
402 shows a broadened distribution with a tail towards the higher delay values, while this tail is greatly reduced for 404.
Due to the shallow depletion zone little charge is collected by drift.
Given a long enough timing window diffusion can drive the signal over threshold, which results in a tail towards higher values.
After irradiation the slowly collected charge is trapped, resulting in a significantly improved timing resolution.
The remaining inaccuracy is attributed to the speed of the amplifier and the response of the discriminator, which has been addressed in the latest prototype versions.
  
\section{Edge TCT results on neutron irradiated \mbox{samples}}
\begin{figure}[htbp]
    \centering
    \includegraphics[width=.95\linewidth]{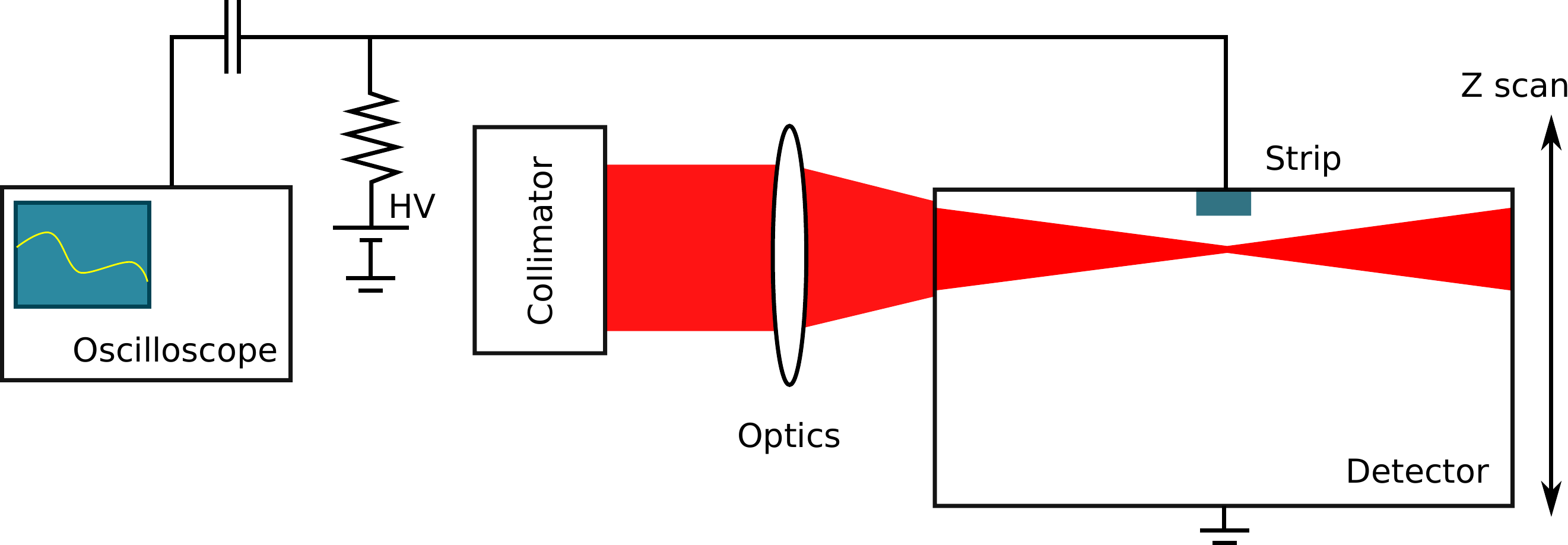}
    \caption{Sketch of a typical Edge TCT setup, consisting of the laser system, fast oscilloscope and biasing electronics}
    \label{fig:etct-setup}
  \end{figure}

The Transient Current Technique is a widely used tool for observing time-resolved charge collection in sensors.
A typically red or infrared laser induces an electrical signal by ionization of the sensor bulk.
Strongly focused lasers provide high spacial resolution and combined with a fast analog readout, movement of charge carriers in the sensor's bulk can be mapped.
Usually, the laser is shot through the top or bottom of the sensor.
In Edge TCT the laser is shot through a polished side of the sensor, allowing to define the charge deposition depth with a precision in the order of micrometers.
A typical experimental setup is sketched in figure \ref{fig:etct-setup}.
\cite{kramberger_etct}
  
Measurements were conducted on H18 CCPDv3 prototypes, which allow direct access to a $100\times100\,\upmu\text{m}^2$ passive diode at the edge of the chips via a wire bonding pad.
It is formed solely by the junction of a deep n-well and the bulk, separated from the fully integrated pixel matrix to avoid influence of the implanted electronics.
Samples used for Edge TCT were irradiated with neutrons at the TRIGA reactor of JSI Ljubljana to $1$, $7$ and \irniel{20}{15}.

\begin{figure}[htbp]
    \centering
    \includegraphics[width=.95\linewidth]{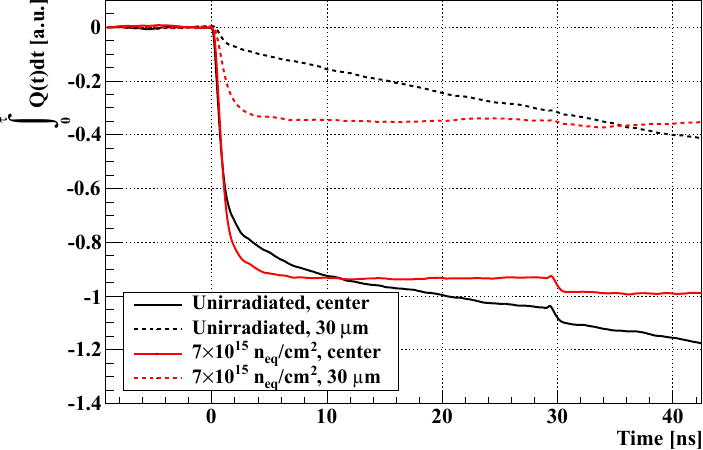}
    \caption{Integrated charge measured in the center of depletion zone and $30\,\upmu$m below for a non irradiated and \irniel{7}{15} irradiated sensor. $U_\text{Bias}=-80\,$V, $T=-20\,^\circ$C \cite{etct_mfgarcia}}
    \label{fig:etct_runningCharge}
  \end{figure}

The analog response to the laser pulses depending on the position and bias voltage was recorded and the charge estimated by integration over the current.
Figure \ref{fig:etct_runningCharge} shows the collected charge integrated over time for a non irradiated and \irniel{7}{15} irradiated sensor.
Charge drifting in the high electric field of the depletion zone will be collected quickly forming a steep edge while diffusing charge is collected slowly creating a tail with a low slope.
This allows to disentangle the drifting from the diffusing charge by applying a timing cut at $t=5\,$ns.
Furthermore, as the induced signal at a depth of $30\,\upmu$m shows only a slow charge collection, the depletion depth for the non irradiated sensor is below $30\,\upmu$m.
In the irradiated case, the signal shows no increase after $5\,$ns, meaning that most of the diffusing current has been trapped and only charge in the depleted zone and the shallow region around, where diffusing charges can reach the depleted zone before being trapped, is collected.
By comparison with the non irradiated data the depletion depth increased after irradiation, which is further investigated by time resolved scans over the depth of the sensor.
  
\begin{figure}[htbp]
    \centering
    \includegraphics[width=.90\linewidth]{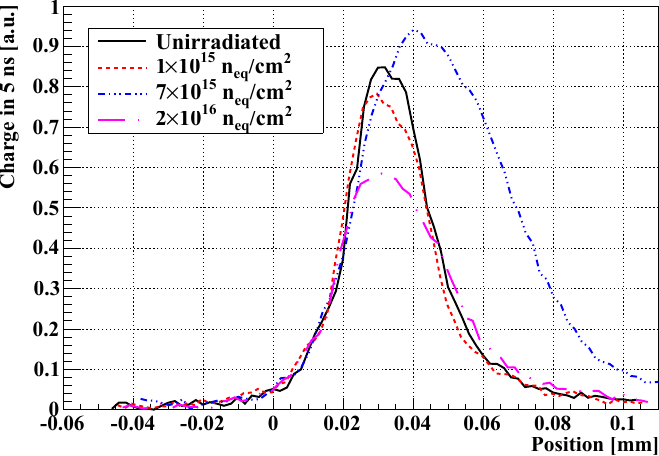}
    \caption{Profile of charge collected in 5\,ns for fluences up to \irniel{2}{16}. $U_\text{Bias}=-80\,$V, $T=-20\,^\circ$C \cite{etct_mfgarcia}}
    \label{fig:etct_chargeProfile}
  \end{figure}

The results of these scans in figure \ref{fig:etct_chargeProfile} reveal the depth profile of the depletion zone.
It is shown, that for \irniel{1}{15} the width is preserved, whereas the amplitude decreases.
After \irniel{7}{15} the profile is significantly widened and also the amplitude is increased by $\approx 10\,\%$, while for \irniel{2}{16} the width is again comparable to the non irradiated case, but here with a significantly reduced amplitude.
 
\begin{figure}[htbp]
    \centering
    \includegraphics[width=.90\linewidth]{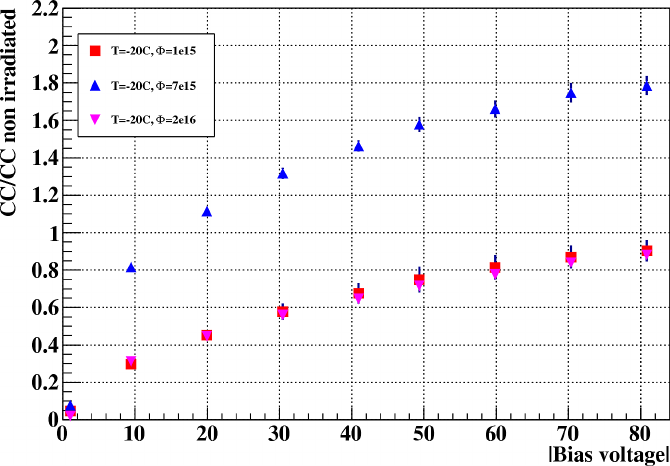}
    \caption{Charge collected in 5\,ns integrated over $90\,\upmu$m depth relatively to a non irradiated sensor versus bias voltage. $T=-20\,^\circ$C \cite{etct_mfgarcia}}
    \label{fig:etct_chargeVsBias}
  \end{figure}

The scan over the depth of the sensor simulates a uniform charge deposition in the bulk like it is caused by MIPs.
By integrating over the charge profile, the collected charge for a MIP can be assessed.
Figure \ref{fig:etct_chargeVsBias} shows this value for different bias voltages.
In the case of unbiased sensors, all irradiated samples collect less than $10\,\%$ of the charge of the unirradiated ones.
The \irniel{1}{15} and \irniel{2}{16} irradiated sensors show a comparable behavior and reach about $90\,\%$ of the charge collection for $80\,$V.
As in the charge profiles, the sensor behaves differently after \irniel{7}{15}.
The charge collection is on par with the non irradiated one already around $15\,$V and eventually almost doubles for $80\,$V.

The influence of radiation on the collected charge can be explained by the acceptor removal effect that occurs in highly doped silicon.
Here the effective doping concentration decreases with fluence, leading to an effectively higher resistive substrate, thus a bigger depletion zone at a given bias voltage.
The competing trapping effect leads to a decreased collection efficiency after \irniel{1}{15}.
In conjunction with the acceptor introduction effect the active zone is narrowed to the non irradiated width after \irniel{2}{16}.

Edge-TCT results on irradiated H35 sensors will be soon published in \cite{etct_h35} and \cite{etct_todd}.
 
\section{The ams H35 Demonstrator}
\begin{figure}[htbp]
    \centering
    \includegraphics[width=.7\linewidth]{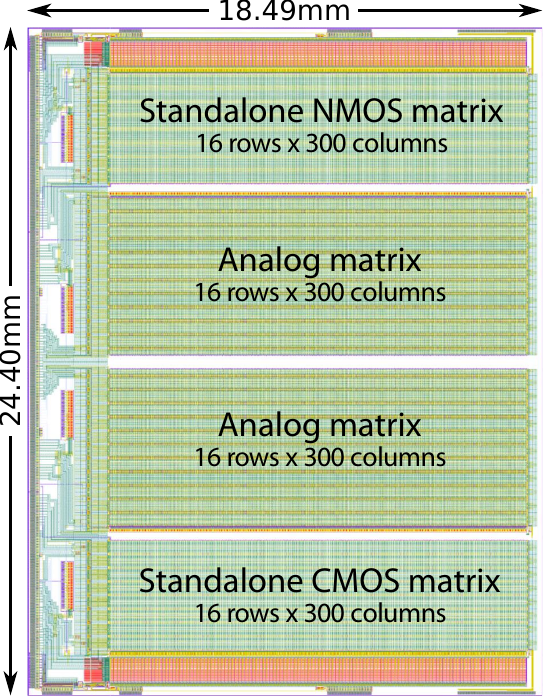}
    \caption{Top view of the demonstrator chip. The four pixel matrices, the separating gap and the standalone readout blocks located at the periphery of the chip are visible.}
    \label{fig:demo_chip}
  \end{figure}

The HV-CMOS sensors characterized so far were small prototypes ($\approx 8\ldots20\,\text{mm}^2$).
Based on the promising characterization results and in order do prove the feasibility of a large scale application, a full reticle size demonstrator chip was designed and submitted through an engineering run.
This $18.5\times24.4\,\text{mm}^2$ sized sensor incorporates four pixel matrices with $250\times50\,\upmu\text{m}^2$ sized pixels, thus being fully compatible to the FE-I4 footprint (see figure \ref{fig:demo_chip}).
The inner two analog matrices are identical, consisting of $23\times300$ pixels containing only amplifiers in three flavors, optimized for gain and/or speed.
These cells have to be read out by a separate readout chip.
The standalone CMOS Matrix consists of $16\times300$ pixel cells holding amplifiers, which can be additionally read out by CMOS comparators in the periphery of the chip.
A digital logic stores the time stamps and generates hit addresses which then can be read out serially at $320\,$MHz.
For the standalone NMOS matrix only NMOS type transistors were implemented. 
The amplified signals are digitized by in-pixel discriminators and only the digital logic remains at the periphery.
All matrices are electrically independent and can be wirebonded and operated separately.
The chip has been submitted on different substrates with resistivities ranging from $20$ to $1000\,\Omega$cm.
For the higher resistivities, bigger depletion zones, thus increased collected charge is expected.
Delivery is envisaged for the end of the year.
  
\section{Summary}
Prototypes of the ams HV-CMOS pixel sensors, designed for the Phase-II upgrade of the ATLAS Pixel Detector, were characterized towards performance and radiation tolerance up to $1\,$Grad TID (X-ray and proton irradiation) and \irniel{2}{16} (thermal neutron irradiation).
The amplifying electronics remained functional to the highest applied doses with a reduced gain, which was partially mitigated by room temperature annealing and retuning of the corresponding DAC parameters.
Effects of NIEL irradiation were assessed in testbeam and Edge TCT measurements.
Although the measured samples could not be biased to the design voltage during the testbeam, the non irradiated as well as the \irniel{1}{15} irradiated sensor showed high detection efficiency of over $99\,\%$, respectively $96\,\%$.
Edge-TCT revealed a strong, drift based signal after irradiation which drives the charge collection between \irniel{1}{15} and \irniel{2}{16} even above the non irradiated case.
Due to the acceptor removal effect, charge collection for \irniel{7}{15} doubles at $80\,$V bias.

The measurements and improvements over the prototype generations lead to the design and submission of a full reticle size demonstrator to prove the large scale application of these sensors for the ATLAS upgrade. 

\section{Acknowledgements}
The research leading to these results has received funding from the European Commission under the FP7 Research Infrastructures project AIDA grant agreement no. 262025.

\section*{References}

\bibliography{mybibfile}

\end{document}